\documentclass[12pt]{article}
\usepackage{amsfonts}
\usepackage{amsmath}
\usepackage{amssymb}
\def\PLB#1#2#3{Phys.\ Lett.\ {\bf B#1} (#3), #2}
\def\NPB#1#2#3{Nucl.\ Phys.\ {\bf B#1} (#3), #2}

\begin{document} 

\begin{flushright}

\end{flushright}

\vspace{3mm}

\begin{center}
{\Large \bf $SU(3)_{\rm L} \rtimes (\mathbb{Z}_3 \times \mathbb{Z}_3)$  
      gauge symmetry 

\vspace{3mm}

      and Tri-bimaximal mixing }

\vspace{15mm}

Chuichiro HATTORI, 
            \footnote{E-mail: hattori@aitech.ac.jp} 
Mamoru MATSUNAGA,$^a$ 
            \footnote{E-mail: matsuna@phen.mie-u.ac.jp} \\
Takeo MATSUOKA$^a$
            \footnote{E-mail: t-matsu@siren.ocn.ne.jp}
and Kenichi NAKANISHI,$^b$ 
            \footnote{deceased}
\end{center}

\begin{center}
\textit{
Science Division, General Education, Aichi Institute of Technology, \\
     Toyota 470-0392, JAPAN \\
{}$^a$Department of Physics Engineering, Mie University, \\
     Tsu 514-8507, JAPAN \\
{}$^b$Department of Sustainable Resource Science, Mie University, \\
     Tsu 514-8507, JAPAN
 }
\end{center}

\vspace{5mm}

\begin{abstract}
We study an  effective gauge theory whose gauge group is 
a semidirect product $G = G_c \rtimes \mathit{\Gamma}$ with $G_c$ 
and $\mathit{\Gamma}$ being a connected Lie group and a finite 
group, respectively. 
The semidirect product is defined through a projective homomorphism 
$\gamma$ (i.e., homomorphism up to the center of $G_c$) from 
$\mathit{\Gamma}$ into $G_c$. 
The (linear) representation of $G$ is made from $\gamma$ and 
a projective representation of $\mathit{\Gamma}$ over $\mathbb{C}$. 
To be specific, we take $SU(3)_L$ as $G_c$ and 
$\mathbb{Z}_3 \times \mathbb{Z}_3$ as $\mathit{\Gamma}$. 
It is noticed that the irreducible projective representations of 
$\mathit{\Gamma}$ are three-dimensional in spite of its Abelian nature. 
We give a toy model on the lepton mixing which illustrates 
the peculiar feature of such gauge symmetry. 
It is shown that under a particular vacuum alignment 
the tri-bimaximal mixing matrix is reproduced. 
\end{abstract}

\newpage 
%%%%%%  SECTION  1  %%%%%%%%%%%%%%%%%%%%%%%%%%%%%%%%%%%%%%%%
\section{Introduction}

Current data on lepton mixing show that the mixing matrix takes 
a peculiar form which is approximately tri-bimaximal.\cite{HPS} 
In order to unify the three families into irreducible flavor 
triplets and to reproduce the specific mixing pattern, 
many authors have searched for various types of non-Abelian 
flavor symmetry.\cite{Flavor} 
A large number of flavor models on lepton mixing 
have been constructed. 
In these attempts the authors have taken up a direct product 
of the flavor group $\mathit{\Gamma}$ with 
the connected gauge group $G_c$. 
However, it is possible to construct models not only with 
direct product groups but also with semidirect product groups. 
We will focus our attention in this paper on a semidirect 
product gauge group $G_c \rtimes \mathit{\Gamma}$.

It has been argued that all non-gauge symmetries are strongly 
violated by quantum gravity effects around the Planck scale and 
hence in the low-energy effective theory we cannot have any global 
symmetries including discrete groups.\cite{Banks} 
For this reason the flavor group $\mathit{\Gamma}$ which appears 
in the low-energy effective theory should be a gauge symmetry. 
Consequently, as for the choice of the gauge symmetry it is 
legitimate to adopt the semidirect product group 
$G_c \rtimes \mathit{\Gamma}$.

In the case of the direct product gauge symmetry 
$G_c \times \mathit{\Gamma}$, 
if the flavor group $\mathit{\Gamma}$ is Abelian, 
it is impossible to have irreducible flavor triplets. 
However, in irreducible linear representations of the semidirect 
product group $G = G_c \rtimes \mathit{\Gamma}$, 
even if $\mathit{\Gamma}$ is Abelian, 
it is possible to have flavor triplets which are 
irreducible projective representations of $\mathit{\Gamma}$. 
In view of this, we study in this paper the effective theory 
with the semidirect product gauge symmetry 
$G = G_c \rtimes \mathit{\Gamma} = 
SU(3)_{\rm L} \rtimes (\mathbb{Z}_3 \times \mathbb{Z}_3)$. 
As an illustrative example, we give a toy model on the lepton 
mixing in the context of the supersymmetric effective theory 
with this gauge symmetry. 
In this model three families of leptons are assigned into flavor 
triplets of the projective representation of 
$\mathit{\Gamma} = \mathbb{Z}_3 \times \mathbb{Z}_3$.

This paper is organized as follows. 
In section 2 we give a brief explanation of the semidirect product 
group $G = G_c \rtimes \mathit{\Gamma}$ and its irreducible 
representations. 
Our emphasis is placed on the projective homomorphism 
(i.e., homomorphism up to the center of $G_c$) of 
$\mathit{\Gamma} = \mathbb{Z}_3 \times \mathbb{Z}_3$ into 
$G_c = SU(3)$. 
In section 3 we present a toy model which illustrates the peculiar 
features of the semidirect product gauge group. 
We show that under a particular vacuum alignment the tri-bimaximal 
lepton mixing can be reproduced. 
Final section is devoted to summary.

\vspace{10mm}

%%%%% SECTION  2 %%%%%%%%%%%%%%%%%%%%%%%%%%%%%%%%%%%%%%%%%%%

\section{Semidirect product group and its representations}

The semidirect product group 
$G = G_c \rtimes \mathit{\Gamma}$ 
which we consider in this paper has the multiplication rule 
\begin{equation}
  (g, \, s) (g', \, s') = (g \, \imath_s(g'), \, ss') 
\end{equation}
for $(g, \, s), (g', \, s')$ with $g, \, g' \in G_c$ 
and $s, \, s' \in \mathit{\Gamma}$. 
The symbol $\imath$ represents a homomorphism from 
$\mathit{\Gamma}$ to the inner automorphism group of $G_c$ 
\begin{equation}
\begin{array}{rcl}
    \imath \, : \ \mathit{\Gamma} & \longrightarrow & \ {\rm Inn}(G_c) \\
                     s   & \longmapsto     & \imath_s(\bullet) 
                                 = \gamma_s (\bullet) \gamma_s^{-1} 
\end{array}
\end{equation}
with $\gamma_s$ being an element of $G_c$.

The homomorphic nature of $\imath$ requires that the product of 
$\gamma_s$ and $\gamma_{s'}$ is equal to $\gamma_{ss'}$ up to 
the center of $G_c$, i.e., 
\begin{equation}
   \gamma_s \, \gamma_{s'} = f_{s,s'} \, \gamma_{ss'} 
                   \quad \left( f_{s,s'} \in Z(G_c) \right), 
\end{equation}
which means that the map 
\begin{equation}
\begin{array}{rcl}
    \gamma \, : \ \mathit{\Gamma} & \longrightarrow & G_c \\
                     s   & \longmapsto     &  \gamma_s \, .
\end{array}
\end{equation}
is a homomorphism up to the center, 
hence called in this paper a {\it projective homomorphism}. 
The set $\{ f_{s,s'} \}_{s,s'\in \mathit{\Gamma}}$ is called 
a factor set associated with $\gamma$. 
The associative law of the products among ${\gamma_s}$'s 
leads to the so-called cocycle condition on the factor set. 
By multiplying  ${\gamma_s}$ by  an  element of $Z(G_c)$, 
we may redefine $\gamma_s$.
If all elements of the factor set are reducible to unity 
through this redefinition, 
it can be shown that the semidirect group is isomorphic to 
the direct product group.\cite{in_prep} 
In other words, in the genuine projective homomorphism 
we can not reduce all of $f_{s,s'}$'s to unity 
via such redefinition. 

We now take $G_c = SU(3)$ and 
\begin{equation}
    \mathit{\Gamma} =\mathbb{Z}_3 \times \mathbb{Z}_3 
       = \langle \, a, \ b \,|\,a^3 = b^3 = e, \ ba = ab \, \rangle. 
\end{equation}
For the projective homomorphism from 
$\mathit{\Gamma} = \mathbb{Z}_3 \times \mathbb{Z}_3$ to $SU(3)$, 
we can set, through the redefinition of  ${\gamma_s}$, 
\begin{equation}
\begin{array}{rcl}
     {\gamma_a}^3 & = & f_a \, {\bf 1}, \\
     {\gamma_b}^3 & = & f_b \, {\bf 1}, \\
     \gamma_b \, \gamma_a & = & f_c \, \gamma_a \, \gamma_b 
\end{array}
\end{equation}
with $f_{a,b,c} {\bf 1} \in Z(SU(3))$.\cite{in_prep} 
In the following we choose $f_a = f_b = 1$ and 
$f_c = \omega = \exp(2\pi i/3)$.

An irreducible {\it linear representation} of 
$G = SU(3) \rtimes \mathit{\Gamma}$ is expressed as 
\begin{equation}
   {\cal R}(g, \, s) = R(g \gamma_s) \otimes \rho_s^* \, ,
\label{eqn:Rgx}
\end{equation}
where $R(g)$ is a representation of $g \in SU(3)$ and 
$\rho_s$ stands for a {\it projective representation} of 
$\mathit{\Gamma}$ over $\mathbb{C}$ whose factor set is 
the same as that of $R(\gamma_s)$. 
In the product of ${\cal R}(g, \, s)$ and ${\cal R}(g', \, s')$, 
the factor sets coming from $R$ and $\rho$ cancell out 
and hence ${\cal R}$ forms a linear representation.

When we take the trivial representation for $R$, 
$\rho$ is one of the irreducible linear representation of 
$\mathbb{Z}_3 \times \mathbb{Z}_3$, which are all singlets 
distinguished by the characters 
\begin{equation}
   ( \chi_a, \ \chi_b ) = ( \omega^k, \ \omega^l ) 
\end{equation}
with $ k, \, l = 0, \ \pm 1$. 
We denote these singlets by ${\bf 1}_{(k,l)}$. 
When we take the fundamental representation ${\bf 3}$ 
(anti-fundamental representation ${\bf 3^*}$) for $R$, 
$\rho$ is the irreducible projective representation  
${\bf 3}$ (${\bf 3^*}$) of $\mathbb{Z}_3 \times \mathbb{Z}_3$, 
which satisfies \cite{Karpilovsky} 
\begin{equation}
\begin{array}{rcl}
     {\rho_a}^3 & = & {\bf 1}, \\
     {\rho_b}^3 & = & {\bf 1}, \\
     \rho_b \, \rho_a & = &  f_c' \, \rho_a \,\rho_b 
\end{array}
\end{equation}
with $f_c' = \omega$ ($\omega^* $). 
The projective unitary representation ${\bf 3}$, 
in a basis where $\rho_a$ is diagonal, is given by 
\begin{equation}
  \rho_a = \left(
           \begin{array}{ccc}
             1  &    0   &     0      \\
             0  & \omega &     0      \\
             0  &    0   &  \omega^2 
           \end{array} 
           \right),   \qquad 
  \rho_b = \left(
           \begin{array}{ccc}
             0  &  0  &  1    \\
             1  &  0  &  0    \\
             0  &  1  &  0 
           \end{array} 
           \right).  
\end{equation}
In this projective representation of $\mathit{\Gamma}$ 
the multiplication rules are 
\begin{eqnarray}
  {\bf 3} \times {\bf 3} \ 
            & = & {\bf 3^*} + {\bf 3^{*'}} + {\bf 3^{*''}}, \\
  {\bf 3} \times {\bf 3^*} 
            & = & \sum_{k,l=0,\pm 1} {\bf 1}_{(k,l)}. 
\end{eqnarray}
The product rules of two triplets $(x_1, x_2, x_3)$ and 
$(y_1, y_2, y_3)$ turn out to be 
\begin{equation}
  \left(
    \begin{array}{c}
      x_1 y_1  \\
      x_2 y_2  \\
      x_3 y_3  
    \end{array}
  \right) \sim {\bf 3^*}, \qquad 
  \left(
    \begin{array}{c}
      x_2 y_3  \\
      x_3 y_1  \\
      x_1 y_2  
    \end{array}
  \right) \sim {\bf 3^{*'}}, \qquad 
  \left(
    \begin{array}{c}
      x_3 y_2  \\
      x_1 y_3  \\
      x_2 y_1  
    \end{array}
  \right) \sim {\bf 3^{*''}}. 
\end{equation}
The product rules of a triplet $(x_1, x_2, x_3)$ and 
an anti-triplet $(\overline{y}_1, \overline{y}_2, \overline{y}_3)$ 
become 
\begin{equation}
\begin{array}{lcl}
  x_1 \overline{y}_1 + x_2 \overline{y}_2 
                        + x_3 \overline{y}_3 & \sim & {\bf 1}_{(0,0)}, \\
  x_1 \overline{y}_1 + \omega^2 x_2 \overline{y}_2 
                 + \omega x_3 \overline{y}_3 & \sim & {\bf 1}_{(0,1)}, \\
  x_1 \overline{y}_1 + \omega x_2 \overline{y}_2 
              + \omega^2 x_3 \overline{y}_3 & \sim & {\bf 1}_{(0,-1)}, \\
  x_1 \overline{y}_3 + x_2 \overline{y}_1 
                        + x_3 \overline{y}_2 & \sim & {\bf 1}_{(1,0)}, \\
  x_1 \overline{y}_3 + \omega^2 x_2 \overline{y}_1 
                 + \omega x_3 \overline{y}_2 & \sim & {\bf 1}_{(1,1)}, \\
  x_1 \overline{y}_3 + \omega x_2 \overline{y}_1 
              + \omega^2 x_3 \overline{y}_2 & \sim & {\bf 1}_{(1,-1)}, \\
  x_1 \overline{y}_2 + x_2 \overline{y}_3 
                       + x_3 \overline{y}_1 & \sim & {\bf 1}_{(-1,0)}, \\
  x_1 \overline{y}_2 + \omega^2 x_2 \overline{y}_3 
                + \omega x_3 \overline{y}_1 & \sim & {\bf 1}_{(-1,1)}, \\
  x_1 \overline{y}_2 + \omega x_2 \overline{y}_3 
             + \omega^2 x_3 \overline{y}_1 & \sim & {\bf 1}_{(-1,-1)}. 
\end{array}
\end{equation}

Generally, the linear representations of 
the semidirect product gauge group 
$SU(p) \rtimes (\mathbb{Z}_p \times \mathbb{Z}_p)$ 
with a prime number $p$ stand in need of the projective 
representation of $\mathit{\Gamma} = \mathbb{Z}_p \times \mathbb{Z}_p$, 
which is designated by the relation 
$\rho_b \, \rho_a = f_c' \, \rho_a \, \rho_b$ with 
$f_c' =\exp (2\pi i m/p),\; m = 0,1,..,p-1$. 
The irreducible representations are $p^2$ singlets 
for $m =0$ (linear representations) and $p$-plet for each 
$m \neq 0$ (genuine projective representation).

It is expected that the theory with the semidirect product gauge 
symmetry contains interesting features different from the one 
with the direct product gauge symmetry. 
To see this concretely, taking 
$G =  SU(3)_L \rtimes (\mathbb{Z}_3 \times \mathbb{Z}_3)$ 
gauge group as an example, 
we formulate a toy model on the lepton mixing.

\vspace{10mm}

%%%%%%  SECTION 3   %%%%%%%%%%%%%%%%%%%%%%%%%%%%%%%%%%%%%%%%
\section{A toy model and tri-bimaximal mixing}

Here we concentrate our attention on the mixing matrix in 
the lepton sector. 
It is assumed that the gauge symmetry $SU(2)_{\rm L}$ is enlarged 
to $SU(3)_{\rm L}$ and that $SU(2)_{\rm L}$-doublet and -singlet 
leptons are assigned into 
$(R, \, \rho) \sim ({\bf 3}, \, {\bf 3})$ as 
\begin{equation}
  \vec{L} = \left(
   \begin {array}{c}
       *           \\
      \nu_{\rm L}  \\
      l^-_{\rm L} 
   \end{array}
  \right),   \qquad \quad 
 \vec{l^c} = \left(
   \begin{array}{c}
     {l_{\rm R}}^c  \\
       *          \\
       *          
   \end{array}
  \right). 
\end{equation}
The fields denoted by asterisks do not affect 
the following discussions. 
We now adopt a supersymmetric context and introduce 
two kinds of Higgs fields, which are 
$(R, \, \rho) \sim ({\bf 3}, \, {\bf 3})$ and 
$({\bf 3^*}, \, {\bf 3^*})$ denoted as 
\begin{equation}
  \vec{\phi} = \left(
    \begin {array}{c}
        *       \\
      \phi^0    \\
      \phi^- 
    \end{array}
   \right),   \qquad \quad 
  \vec{\varphi} = \left(
   \begin{array}{c}
      \varphi^+  \\
      \varphi^0  \\
        *    
   \end{array}
  \right), 
\end{equation}
respectively. 
Incidentally, $\vec{L}$ and $\vec{l^c}$ represent odd R-parity 
superfields, 
while $\vec{\phi}$ and $\vec{\varphi}$ even R-parity ones. 
The product rules among flavor triplets and anti-triplets such as 
$(\vec{L_1}, \ \vec{L_2}, \, \vec{L_3})$, 
$(\vec{l^c_1}, \ \vec{l^c_2}, \, \vec{l^c_3})$, etc.\ 
are given in the previous section.

We next proceed to study the $G$-invariant operators in 
the superpotential. 
Taking account of the fact that the element 
$(g, \, s) \in G = G_c \rtimes \mathit{\Gamma}$ is decomposed as 
$(g, \, s) = (g, \, e)(e, \, s)$ uniquely, 
we can replace the $G$-invariant condition on the operators 
by the separate requirements of the $G_c$-invariance 
and $\mathit{\Gamma}$-invariance. 
Charged lepton masses arise from the $SU(3)_{\rm L}$-invariant 
Yukawa couplings $\vec{L} \, \vec{l^c} \, \vec{\phi}$ in 
the superpotential. 
The flavor structure of the Yukawa couplings 
$\vec{L} \, \vec{l^c} \, \vec{\phi}$ is of the form 
\begin{eqnarray}
    h_1 \left( \vec{L_1} \vec{l^c_1} \vec{\phi_1} 
                         + \vec{L_2} \vec{l^c_2} \vec{\phi_2} 
                         + \vec{L_3} \vec{l^c_3} \vec{\phi_3} \right) 
  + h_2 \left( \vec{L_1} \vec{l^c_2} \vec{\phi_3} 
                         + \vec{L_2} \vec{l^c_3} \vec{\phi_1} 
                         + \vec{L_3} \vec{l^c_1} \vec{\phi_2} \right) 
                                   \nonumber \\
  + \, h_3 \left( \vec{L_1} \vec{l^c_3} \vec{\phi_2} 
                         + \vec{L_2} \vec{l^c_1} \vec{\phi_3} 
                         + \vec{L_3} \vec{l^c_2} \vec{\phi_1} \right). \qquad 
\end{eqnarray}
When the Higgs fields $\phi_i^0$ $(i = 1,2,3)$ acquire their vacuum 
expectation values $v_i$, we obtain the charged lepton mass matrix 
\begin{equation}
  M_l = \left( 
         \begin{array}{ccc}
           h_1 v_1  &  h_2 v_3  &  h_3 v_2  \\
           h_3 v_3  &  h_1 v_2  &  h_2 v_1  \\
           h_2 v_2  &  h_3 v_1  &  h_1 v_3  
         \end{array}
        \right). 
\end{equation}
If we assume the vacuum configuration 
\begin{equation}
   v_1 = v_2 = v_3, 
\end{equation}
the charged lepton mass matrix $M_l$ is diagonalized through 
the unitary transformation $U_l^{\dag} M_l U_l$. 
The unitary matrix $U_l$ becomes 
\begin{equation}
  U_l = \frac{1}{\sqrt{3}} \left( 
         \begin{array}{ccc}
           1  &      1     &     1      \\
           1  &   \omega   &  \omega^2  \\
           1  &  \omega^2  &   \omega  
         \end{array}
        \right). 
\label{eqn:Ul}
\end{equation}

Light neutrino masses are described by the non-renormalizable 
superpotential term 
$\vec{L} \, \vec{L} \, \vec{\varphi} \, \vec{\varphi}$, 
where $\vec{L}$ and $\vec{\varphi}$ is assigned into 
$(R, \, \rho) \sim ({\bf 3}, \, {\bf 3})$ and 
$({\bf 3^*}, \, {\bf 3^*})$, respectively. 
There are two types of the non-renormalizable operators~\cite{Morisi} 
\begin{equation}
   (\vec{L} \, \vec{L})_{\bf 3^*} \, 
                  (\vec{\varphi} \, \vec{\varphi})_{\bf 3}, \qquad \quad 
   (\vec{L} \, \vec{\varphi})_{\bf 1} \, (\vec{L} \, \vec{\varphi})_{\bf 1}. 
\end{equation}
The first type leads to the couplings 
\begin{eqnarray}
  \frac{g_1}{\Lambda} \left( 
             \nu_1 \, \nu_1 \, \varphi_1 \, \varphi_1 
           + \nu_2 \, \nu_2 \, \varphi_2 \, \varphi_2 
           + \nu_3 \, \nu_3 \, \varphi_3 \, \varphi_3  \right) 
                                             \qquad \qquad \nonumber \\
   \qquad \qquad + \, 2 \, \frac{g_2}{\Lambda} \left( 
             \nu_2 \, \nu_3 \, \varphi_2 \, \varphi_3 
           + \nu_3 \, \nu_1 \, \varphi_3 \, \varphi_1 
           + \nu_1 \, \nu_2 \, \varphi_1 \, \varphi_2  \right), 
\end{eqnarray}
where $\Lambda$ is the cutoff and $\nu_i$ and $\varphi_i$ represent 
the component $(\nu_L)_i$ in $\vec{L_i}$ and $\varphi_i^0$ 
in $\vec{\varphi_i}$, respectively. 
When the fields $\varphi_i^0$ have the vacuum expectation values 
$u_i$ $(i = 1,2,3)$, 
the above terms induce the mass matrix term 
\begin{equation}
  M_{\nu}^{(a)} = \frac{1}{\Lambda} \left( 
         \begin{array}{ccc}
           g_1 u_1^{\ 2} &   g_2 u_1 u_2   &  g_2 u_1 u_3  \\
           g_2 u_1 u_2   &   g_1 u_2^{\ 2} &  g_2 u_2 u_3  \\
           g_2 u_1 u_3   &   g_2 u_2 u_3   &  g_1 u_3^{\ 2}  
         \end{array}
        \right) 
\end{equation}
for light neutrinos. 
In the second type, the $\mathit{\Gamma}$-invariant combinations 
of two singlets $(\vec{L} \, \vec{\varphi})_{\bf 1}$ are 
${\bf 1}_{(k,l)} \times {\bf 1}_{(-k,-l)}$ with $k, \, l = 0, \ \pm 1$. 
Thus the second type operators yields the couplings 
\begin{eqnarray}
 & &  \frac{f_1}{\Lambda} \left( \nu_1 \, \varphi_1 
        + \nu_2 \, \varphi_2 + \nu_3 \, \varphi_3 \right)^2 \nonumber \\
 & & + \, 2 \, \frac{f_2}{\Lambda} \left( \nu_1 \, \varphi_1 
        + \omega^2 \nu_2 \, \varphi_2 + \omega \nu_3 \, \varphi_3 \right) 
                 \left( \nu_1 \, \varphi_1 + \omega \nu_2 \, \varphi_2 
                      + \omega^2 \nu_3 \, \varphi_3 \right) \nonumber \\
 & & + \, 2 \, \frac{f_3}{\Lambda} \left( \nu_1 \, \varphi_3 
        + \nu_2 \, \varphi_1 + \nu_3 \, \varphi_2 \right) 
                 \left( \nu_1 \, \varphi_2 + \nu_2 \, \varphi_3 
                      + \nu_3 \, \varphi_1 \right)   \nonumber \\
 & & + \, 2 \, \frac{f_4}{\Lambda} \left( \nu_1 \, \varphi_3 
        + \omega^2 \nu_2 \, \varphi_1 + \omega \nu_3 \, \varphi_2 \right) 
                 \left( \nu_1 \, \varphi_2 + \omega \nu_2 \, \varphi_3 
                    + \omega^2 \nu_3 \, \varphi_1 \right)   \nonumber \\
 & & + \, 2 \, \frac{f_5}{\Lambda} \left( \nu_1 \, \varphi_3 
        + \omega \nu_2 \, \varphi_1 + \omega^2 \nu_3 \, \varphi_2 \right) 
                 \left( \nu_1 \, \varphi_2 + \omega^2 \nu_2 \, \varphi_3 
                    + \omega \nu_3 \, \varphi_1 \right). 
\end{eqnarray}
Taking $f_3 = f_4 = f_5$, we have another mass matrix term 
\begin{eqnarray}
  M_{\nu}^{(b)} = \frac{1}{\Lambda} \left( 
         \begin{array}{ccc}
           (f_1+2f_2)u_1^{\ 2} &  (f_1-f_2) u_1 u_2  &  (f_1-f_2) u_1 u_3  \\
           (f_1-f_2) u_1 u_2  &  (f_1+2f_2)u_2^{\ 2} &  (f_1-f_2) u_2 u_3  \\
           (f_1-f_2) u_1 u_3  &  (f_1-f_2) u_2 u_3  &  (f_1+2f_2) u_3^{\ 2}  
         \end{array}
        \right)   \nonumber  \\
    + \frac{1}{\Lambda} \left( 
         \begin{array}{ccc}
           6f_3 u_2 u_3   &       0         &       0        \\
                 0        &  6f_3 u_3 u_1   &       0        \\
                 0        &       0         &  6f_3 u_1 u_2  
         \end{array}
        \right). 
\end{eqnarray}
We now assume a particular vacuum alignment $u_1 = 0$ and $u_2 = u_3$. 
In this case the light neutrino mass matrix becomes 
\begin{eqnarray}
  M_{\nu} & = & M_{\nu}^{(a)} + M_{\nu}^{(b)}  \nonumber  \\
          & = & \frac{u_2^{\ 2}}{\Lambda} \left( 
         \begin{array}{ccc}
           6f_3   &        0         &        0         \\
            0     &  (g_1+f_1+2f_2)  &  (g_2+f_1-f_2)   \\
            0     &  (g_2+f_1-f_2)   &  (g_1+f_1+2f_2)   
         \end{array}
        \right). 
\end{eqnarray}
This mass matrix is diagonalized through the transformation 
$U_{\nu}^T M_{\nu} U_{\nu}$. 
The unitary matrix $U_{\nu}$ is of the form 
\begin{equation}
  U_{\nu} = \left( 
         \begin{array}{ccc}
                  0           &   1   &         0            \\
           \frac{1}{\sqrt 2}  &   0   &  \frac{-i}{\sqrt 2}  \\
           \frac{1}{\sqrt 2}  &   0   &  \frac{i}{\sqrt 2}
         \end{array}
        \right). 
\label{eqn:Unu}
\end{equation}

The lepton mixing matrix is given by $U_{l\nu} = U_l^{\dag} U_{\nu}$. 
Referring to Eqs.~(\ref{eqn:Ul}) and (\ref{eqn:Unu}), 
we find that the lepton mixing is tri-bimaximal, i.e., 
\begin{equation}
   U_{l\nu} =   \left( 
      \begin{array}{ccc}
         {\sqrt \frac{2}{3}}  &  \frac{1}{\sqrt 3}  &         0           \\
         - \frac{1}{\sqrt 6}  &  \frac{1}{\sqrt 3}  & - \frac{1}{\sqrt 2} \\
         - \frac{1}{\sqrt 6}  &  \frac{1}{\sqrt 3}  &   \frac{1}{\sqrt 2}
      \end{array}
      \right). 
\label{eqn:UTB}
\end{equation}

\vspace{10mm}

%%%%%%  SECTION 4   %%%%%%%%%%%%%%%%%%%%%%%%%%%%%%%%%%%%%%
\section{Summary}
In this paper we have studied the effective theory with the 
semidirect product gauge group 
$G = SU(3)_{\rm L} \rtimes (\mathbb{Z}_3 \times \mathbb{Z}_3)$. 
In the semidirect product gauge group 
we need to introduce the projective homomorphism of 
$\mathbb{Z}_3 \times \mathbb{Z}_3$ into $SU(3)_{\rm L}$. 
In spite of Abelian nature of $\mathbb{Z}_3 \times \mathbb{Z}_3$, 
there appear flavor triplets. 
This result is general features of the semidirect product gauge group. 
We applied this result to the issue of the lepton mixing and 
gave a toy model. 
It was shown that under a particular vacuum alignment 
the tri-bimaximal mixing matrix is reproduced.

It is feasible that the semidirect product gauge group 
$G = G_c \rtimes \mathit{\Gamma}$ is traced back to a unified gauge group 
${\tilde G}$. 
Namely, there is a possibility that the symmetry breaking of ${\tilde G}$ 
down to $G = G_c \rtimes \mathit{\Gamma}$ takes place via 
the Higgs mechanism. 
A simple example illustrative of such breaking structure is 
the symmetry breaking $SO(3) \rightarrow SO(2) \rtimes \mathbb{Z}_2$ 
which has been discussed by Preskill and Krauss.\cite{Preskill} 
The theories with the semidirect product gauge group have more 
attractive and rich structure compared to those with the direct 
product gauge group. 
It is expected that the study of theories with the semidirect 
product gauge group sheds new light on the development of 
gauge theories.

\vspace{10mm}

%%%%%  ACKNOWLEDGEMENTS  %%%%%%%%%%%%%%%%%%%%%%%%%%%%%%%%%%%
\section*{Acknowledgements}
One of the authors (M. M.) is grateful to Jyun'ichi Iwamoto 
and Shoushirou Sugio for discussions at the early stage of 
this research.

\vspace{5mm}

%%%%%  REFERENCES  %%%%%%%%%%%%%%%%%%%%%%%%%%%%%%%%%%%%%%%%%

%%% END %%%%%%%%%%%%%%%%%%%%%%%%%%%%%%%%%%%%%%%%%%

\end{document}